\documentclass[a4paper,twocolumn,11pt,accepted=2020-06-08]{quantumarticle}
\pdfoutput=1
\usepackage[utf8]{inputenc}
\usepackage[english]{babel}
\usepackage[T1]{fontenc}
\usepackage{amsmath}
\usepackage{hyperref}

\usepackage{tikz}
\usepackage{lipsum}
\usepackage{csquotes}
\usepackage[final]{changes}
\usepackage[mathlines]{lineno}
\usepackage{cite}

\begin{document}

\title{Switching Quantum Reference Frames for Quantum Measurement}

\author{Jianhao M. Yang}
\email{jianhao.yang@alumni.utoronto.ca}
\affiliation{Qualcomm, San Diego, CA 92121, USA}
\orcid{0000-0002-2828-3618}

\maketitle

\begin{abstract}
Physical observation is made relative to a reference frame. A reference frame is essentially a quantum system given the universal validity of quantum mechanics. Thus, a quantum system must be described relative to a quantum reference frame (QRF). Further requirements on QRF include using only relational observables and not assuming the existence of external reference frame. To address these requirements, two approaches are proposed in the literature. The first one is an operational approach (F. Giacomini, et al, Nat. Comm. 10:494, 2019) which focuses on the quantization of transformation between QRFs. The second approach attempts to derive the quantum transformation between QRFs from first principles (A. Vanrietvelde, et al, \textit{Quantum} 4:225, 2020). Such first principle approach describes physical systems as symmetry induced constrained Hamiltonian systems. The Dirac quantization of such systems before removing redundancy is interpreted as perspective-neutral description. Then, a systematic redundancy reduction procedure is introduced to derive description from perspective of a QRF. The first principle approach recovers some of the results from the operational approach, but not yet include an important part of a quantum theory - the measurement theory. This paper is intended to bridge the gap. We show that the von Neumann quantum measurement theory can be embedded into the perspective-neutral framework. This allows us to successfully recover the results found in the operational approach, with the advantage that the transformation operator can be derived from the first principle. In addition, the formulation presented here reveals several interesting conceptual insights. For instance, the projection operation in measurement needs to be performed after redundancy reduction, and the projection operator must be transformed accordingly when switching QRFs. These results represent one step forward in understanding how quantum measurement should be formulated when the reference frame is also a quantum system. 
\end{abstract}

\section{Introduction}
\label{intro}
The idea that a physical system or a physical phenomenon must be described relative to a reference frame is a well-accepted principle in the relativity theory. Abandoning the concept of absolute spacetime is a foundation of the general relativity where the laws of physics \replaced{are}{is} invariant when changing reference systems. A reference frame essentially is also a quantum system, if we agree that quantum mechanics is universally valid.  Thus, a physical system or a physical phenomenon must be described relative to another quantum system. This statement is not only applied to describe a relativity event, but also applicable to descriptions of all quantum phenomena. The implication here is that a more fundamental theory should describe a physical system relative to a quantum reference frame (QRF), and address how such descriptions are transformed from one to another when switching the QRFs. 

There are extensive research literature related to QRFs~\cite{QRF1, QRF2, QRF3, QRF4, QRF5, QRF6, QRF7, QRF8, QRF9, QRF10, QRF11, QRF12, QRF13, QRF14, QRF15, QRF16}. Here we are only interested to those theories that satisfy two criteria: 1.) Completely abandoning the concept of classical system. All reference systems are quantum systems instead of some kinds of abstract entities. Treating a reference frame as a classical system, such as how the relativity theory does, should be considered as an approximation of a more fundamental theory that is based on QRF. 2.) Completely abandoning any external reference system and the concept of absolute state. Physical description is constructed using relational variables from the very beginning. These criteria are at the heart of the relational quantum mechanics (RQM)~\cite{Rovelli96, Rovelli07, Transs2018, Rovelli18, Yang2017, Yang2018, Hoehn2014, Hoehn2015}\footnote{In the context of relational quantum mechanics~\cite{Rovelli96, Rovelli07, Transs2018, Rovelli18}, a quantum system needs to be described relative to another quantum system~\cite{Rovelli96}. RQM discards the separation of classical system and quantum system and assumes all systems are quantum systems, including macroscopic systems. The relational properties between two systems are more basic than the independent properties of a system. An implementation of RQM is constructed such that quantum mechanics can be reformulated with relational properties as starting point~\cite{Yang2017, Yang2018}.}. \added{We may call approaches based on these two criteria as RQM approaches. This second criteria is crucial to differentiate the RQM approaches from other important research literature on QRF, especially the resource theory~\cite{QRF14, QRF15}. A comprehensive review of difference between resource theory and the RQM approaches is provided in Ref.~\cite{Brukner}.} 

Given these criteria, two RQM approaches stand out as of importance. The first one is an operational approach~\cite{Brukner} that focuses on how the dynamical physical laws are transformed when switching QRFs. By operational approach, we mean it assumes that every QRF is equipped with hypothetical measuring apparatus that can perform measurement to justify a state assignment. The theory has produced several interesting results. For instance, it allows us to derive the operational meaning of spin of a relativistic particle; It shows that entanglement among quantum systems depends on the choice of QRF; It also shows how the measurement outcomes from perspectives of two different QRFs can be related. The limitation of the operational approach is that it assumes the transformation between two QRFs is known beforehand. The operator associated with the transformation is derived primarily from intuition. 

The second approach~\cite{Hoehn2018}, which we call it a first principle approach, attempts to address the limitation in the operational approach. The starting point in this approach is the same, that is, quantum systems should be described with relational variables only. What is novel in this approach is that it implements such idea using the tools and concepts of constraint Hamilton systems~\cite{Dirac, Henneaux}\footnote{The theory of constraint Hamilton system is also key in the canonical formulation of general relativity and quantum gravity~\cite{Rovelli2004, Thiemann}. Thus, the first principle approach inherits such advantage and is potentially an important step towards a successful quantum gravity theory.}. The constraint conditions encode the corresponding gauge symmetries. For instance, the translational symmetry results in a constraint that total momentum to be zero. This implies the system should be described in a constraint surface in the phase space. There are two commonly used methods to canonically quantize the constrained systems, namely, the reduced quantization, and the Dirac quantization. In the reduced quantization method, one solves the constraints first in the context of classical mechanics, then quantizes the reduced theory. Choosing a perspective of a particular reference frame is interpreted as imposing a gauge fixing condition. Thus this method derives the quantum theory directly for a particular quantum reference frame. On the other hand, in the Dirac quantization method, one quantizes the system first without considering the constraints, then solves the constraints in the quantum theory. The quantized theory before removing the symmetry related redundancy is interpreted as a perspective-neutral theory and it does not admit immediate operational interpretation. It is essentially a global description of physics prior to having chosen a reference frame relative to which the physics is described. When the redundancy is removed, the theory is reduced to the perspective from a particular QRF. The reduced theory then admits an operation interpretation. Removing the symmetry is equivalent to fixing the gauge in the case of classical mechanics. However, there is a subtlety that in some generic systems such as the N-body three dimensional systems, there is no globally valid gauge fixing condition~\cite{Hoehn2018-2}. Regardless of the interpretation, the essence of the procedure is to remove the symmetry related redundancy after the Dirac quantization. The procedure confirms that the Dirac quantization theory can be consistently mapped to the reduced quantization theory in the simple one-dimensional toy model. The procedure can be applied to different QRFs, resulting reduced quantum theory for different QRFs, this allows us to derive the transformation formulation between QRFs.

The first principle approach has successfully recovered some of the results in the operational approach, with the advantage that the transformation operators are naturally derived. These include the quantum state transformation when switching QRFs, the Schr\"{o}dinger equation, and the conclusion that entanglement among quantum systems depends on the chosen QRF. However, an important part of a quantum theory, the measurement theory, is not yet studied. 

The goal of this paper is to investigate how the measurement theory can be embedded into the first-principle approach and how to recover the results related to measurements in the operational approach. We focus on the von Neumann quantum measurement theory~\cite{Neumann}, where the measurement stages are separated into two sub processes, namely, Process 1 and Process 2. In Process 2, the measured system and the measuring apparatus interact, but the combined systems can be described as unitary process and determined by the Schrodinger Equation. At the end of the interaction, the combined systems are entangled, and multiple outcomes are possible. Each of the outcome is associated with a specific value of a pointer variable of the apparatus and a probability. Then, in Process 1, which is described by a projection operation, a specific outcome is singled out. \deleted{We will show that Process 2 can be described in the perspective-neutral framework, while Process 1 in general needs to be described with respect to a particular QRF. When the pointer variable is invariant in the symmetry reduction procedure, it is mathematically equivalent to describe Process 1 in the perspective-neutral framework as well.}  In Section \ref{measurement}, we show that the von Neumann measurement theory can be successfully embedded into the perspective-neutral framework. We are able to derive the measurement formulations relative to different QRFs and the transformation between them. The results are consistent with the results in the operational approach~\cite{Brukner} but with the advantage that they are based on a generic first principle approach. Thus the results can be applied to more complex systems such as three-dimensional many-body system. 

\replaced{Although the works presented here extensively uses the theory of the first principle approach}{Furthermore}, our results reveal \replaced{novel }{additional} insights on the measurement theory when switching QRFs \added{that are not presented in Ref.~\cite{Hoehn2018}}. 1.) \replaced{At the concept level, we show that Process 2 can be described in the perspective-neutral framework. However, Process 1 is perspective and in general needs to be implemented after a QRF is chosen, except in the special case where the pointer variable is invariant in the symmetry reduction procedure.}{It is crucial to consider the order of operations when formulating the measurement theory: should we perform the quantum symmetry reduction procedure first, or projection first?} \added{2.) At the methodology level, we show how the symmetry reduction procedure can be embedded in the unitary formulation of Process 2, and then consistently integrated with projective operation in the reduced Hilbert space. There are ample calculation examples in Section \ref{measurement} and the appendix to demonstrate this technique. 3.) At the application level, we discuss in Section \ref{discussion} the importance of synchronizing the projection operator when switching QRFs, which further confirms the synchronization principle proposed in Ref.~\cite{Yang2018, Yang2019} to resolve the extended Wigner's friend paradox. The projection operator needs to be transformed properly in order to preserve the measurement probability when switching QRFs.}

In summary, \replaced{inspired by the works started in}{together with} Ref.~\cite{Brukner, Hoehn2018}, we \replaced{make one step forward}{extend} in the understanding of how quantum \replaced{measurement}{mechanics} is formulated when the reference frame itself is a quantum system and when switching QRFs.

\section{Switching QRFs via a Perspective-Neutral Frameworks}
\label{review}
This section briefly reviews the first principle approach developed in Ref.~\cite{Hoehn2018} for switching QRFs via a perspective-neutral framework. For convenience, we will adopt the same notations used in Ref.~\cite{Hoehn2018}. At the core of the theory is the notion that physics is purely relational. Physically meaningful variables are relational observable, they are invariant under certain gauge transformation. To achieve such description, the systems under \replaced{study}{studied} are described as constrained Hamiltonian systems~\cite{Dirac, Henneaux}. 

\subsection{A Toy Model}
The theory is illustrated through a simple one dimensional toy model. We start to describe the toy model in the context of classical mechanics. The model consists N particles with unit mass and canonical pairs $\{p_i, q_i\}$. These canonical pairs define a $2N$ phase space. To ensure only relational observables are used to describe the systems, the Lagrangian of such systems, in the context of classical mechanics, must be invariant under global translation. This requirement leads to the following constraint
\begin{equation}
\label{constraint}
    P = \sum_i^N p_i \approx 0
\end{equation}
i.e., the momenta of the particles are not all independent. The constraint reduces the phase space dimension to be $2N-1$. The above equation thus defines a constraint surface in the original phase space such that it only holds in that constraint surface (the symbol $\approx$ is used to reflect this weak equality). A Dirac observable $O$ is defined as a variable that is Poisson-commuting with $P$ in the constraint surface, i.e., $\{O, P\}\approx 0$. Obviously all momenta $p_i$ and the relative distance $q_i - q_j$ are all Dirac observables. The total Hamiltonian is read as~\cite{Hoehn2018}
\begin{equation}
    H_{tot} = \frac{1}{2}\sum_i^N p_i^2 + V(\{q_i - q_j\}) + \lambda P
\end{equation}
where $\lambda$ is the Lagrangian multiplier. It is determined when choosing a reference frame thus fixing the gauge. For instance, if one chooses particle $A$ as reference frame, the gauge is fixed, and it may be defined as choosing $q_A = 0$ in the constraint surface. The momentum of $A$ is solved from (\ref{constraint}):
\begin{equation}
\label{constraint2}
    p_A\approx -\sum_{i\ne A}p_i.
\end{equation}
From the equations of motion and the gauge fixing condition $q_A = 0$, one can derive that $\lambda = -p_A$. Then, the corresponding reduced Hamilton from $A$ perspective can be computed as
\begin{equation}
\label{H}
    H_{A}^{red} = \sum_{i\ne A} p_i^2 + \sum_{i \ne j; i,j\ne A} p_ip_j + V(\{q_i\}_{i\ne A}).
\end{equation}

Now we quantize this toy model. For simplicity, only three particles A, B, C, are considered, i.e., $N=3$. First we consider the reduced quantization method. Supposed A is chosen as QRF, we need to quantize the theory in the reduced phase space from A's perspective. The standard procedure is to promote $\{p_i, q_i, i\ne A\}$ to operators that satisfy the following commutation relations:
\begin{equation}
    \label{commutator}
    [\hat{q}_i,\hat{p}_j]=\delta_{ij}, \quad [\hat{q}_i,\hat{q}_j]=[\hat{p}_i,\hat{p}_j]=0, \quad i,j\ne A.
\end{equation}
This defines a $2(N-1)$ dimensional \added{phase space in an infinite} Hilbert space. In the $N=3$ case, the reduced Hamiltonian (\ref{H}) is quantized as
\begin{equation}
\label{H2}
    \hat{H}_{BC|A} = \hat{p}^2_B + \hat{p}^2_C + \hat{p}_B\hat{p}_B + V(\hat{q}_B, \hat{q}_B).
\end{equation}
An arbitrary state vector for particle B and C, from A's perspective, can be written as
\begin{equation}
\label{state1}
    |\psi\rangle_{BC|A} = \int dp_Bdp_c \psi_{BC|A}(p_B, p_C) |p_B\rangle|p_C\rangle.
\end{equation}

\subsection{Dirac Quantization}
In the Dirac Quantization method, one first quantizes the system without considering the constraints. This is achieved by promoting all $\{p_i, q_i\}$ to operators and with appropriate commutation relations. This defines a $2N$ dimensional \added{phase space in an infinite} kinematical Hilbert space, ${\cal{H}}^{kin}$. Next, the momentum constraint (\ref{constraint}) is quantized in this Hilbert space. In the Dirac quantization method, this is amounted to require that the physical states of the systems are annihilated by the momentum constraint operator, i.e.,
\begin{equation}
\label{constraint3}
    \hat{P}|\psi\rangle^{phys} = (\hat{p}_A + \hat{p}_B + \hat{p}_C)|\psi\rangle^{phys} = 0.
\end{equation}
To ensure proper inner product is well-defined for the physical state, a physical Hilbert space ${\cal{H}}^{phys}$ is constructed from ${\cal{H}}^{kin}$ by an improper projector $\delta(\hat{P}): {\cal{H}}^{kin} \to {\cal{H}}^{phys}$. This is defined by the following map\cite{Thiemann},
\begin{equation}
    \label{map}
    |\psi\rangle^{phys} = \delta(\hat{P})|\psi\rangle^{kin} = (\frac{1}{2\pi}\int ds e^{is\hat{P}})|\psi\rangle^{kin}.
\end{equation}
With this definition, from an arbitrary state in ${\cal{H}}^{kin}$,
\begin{equation}
\begin{split}
    |\psi\rangle^{kin} &= \int dp_Adp_Bdp_C\psi^{kin}(p_A,p_B,p_C)\\
    & \times|p_A\rangle|p_B\rangle|p_C\rangle,
\end{split}
\end{equation}
one can obtain the solution for $|\psi\rangle^{phys}$, if solving the constraint for particle A, as
\begin{equation}
\label{phystate}
\begin{split}
    |\psi\rangle^{phys} &= \int dp_Bdp_C\psi_{BC|A}(p_B,p_C)\\
    &\times|-p_B-p_C\rangle_A|p_B\rangle_B|p_C\rangle_C,
\end{split}
\end{equation}
where $\psi_{BC|A}(p_B,p_C)=\psi^{kin}(-p_B-p_C,p_B,p_C)$. It can be verified that (\ref{constraint3}) is satisfied. The sought-after proper inner product for $|\psi\rangle^{phys}$ is
\begin{equation}
    \label{inner}
    (\psi^{phys}, \phi^{phys}) = \langle\psi |\delta(\hat{P})|\phi\rangle^{kin}.
\end{equation}

So far the quantization procedure follows the standard Dirac quantization for constraint system. The novelty in Ref~\cite{Hoehn2018} is to interpret the results of the Dirac quantization as a perspective-neutral framework for the systems. It takes all perspectives of reference systems at once, be it from particle A, or B, or C. To recover the relative states from Dirac quantization to a reduced quantum theory for a specific reference frame, Ref~\cite{Hoehn2018} proposes the following procedure:

Step 1, pick a reference system, say, particle A. The physical state is then given in (\ref{phystate}).

Step 2, apply a unitary transformation defined in  ${\cal{H}}^{kin}$
\begin{equation}
    \label{opT}
    \hat{T}_{A,BC}=e^{i\hat{q}_A(\hat{p}_B+\hat{p}_C)}.
\end{equation}
This effectively defines a new representation of the physical Hilbert space. The physical state (\ref{phystate}) becomes
\begin{equation}
    \label{phystate2}
    \begin{split}
    &|\psi\rangle_{A,BC} := \hat{T}_{A,BC}|\psi\rangle^{phys} \\
    &= |p=0\rangle_A\otimes (\int dp_Bdp_C\psi(p_B,p_C)|p_B\rangle_B|p_C\rangle_C).
    \end{split}
\end{equation}
It is important to note that such transformation is a global operation acting on all three particle at once instead of just a local operation acting on only one particle. Thus, it can change the entanglement properties among the particles. 

Step 3, apply a projection to remove the redundancy of the degree of freedom from the reference system. This also eliminates the self-reference problem. This step is equivalent to project the reference system onto the classical gauge fixing condition. The reduced state from reference system A, is
\begin{equation}
    \label{redustateA}
    \begin{split}
    &|\psi\rangle_{BC|A} = \sqrt{2\pi}\langle q_A=0|\psi\rangle_{A,BC} \\
    &= \int dp_Bdp_C\psi_{BC|A}(p_B,p_C)|p_B\rangle_B|p_C\rangle_C.
    \end{split}
\end{equation}
This recovers the same reduced state as in (\ref{state1}). The resulting Hilbert space is denoted as ${\cal{H}}_{BC|A}$.

\subsection{Switching QRFs}
If we repeat the three steps in the previous subsection but picking particle C as the reference system, we obtain the reduced state
\begin{equation}
    \label{redustateC}
    |\psi\rangle_{AB|C} 
    = \int dp_Adp_B\psi_{AB|C}(p_A,p_B)|p_A\rangle_A|p_B\rangle_B,
\end{equation}
where $\psi_{AB|C}(p_A,p_B)=\psi^{kin}(p_A,p_B,-p_A-p_B)$, and the resulting Hilbert space is denoted as ${\cal{H}}_{AB|C}$. Switching QRF from particle A to C is described by the following map:
\begin{equation}
    \label{map}
    \hat{S}_{A\to C} : {\cal{H}}_{BC|A} \to {\cal{H}}_{AB|C}.
\end{equation}
Such map can be derived by inverting the transformation from A-perspective back to the neutral-perspective framework and then applying the transformation to C-perspective. It connects the two reduced states as $\hat{S}_{A\to C}|\psi\rangle_{BC|A}=|\psi\rangle_{AB|C}$. Ref~\cite{Hoehn2018} shows that this map is equivalent to
\begin{equation}
    \label{map2}
    \hat{S}_{A\to C} = \hat{\cal{P}}_{CA}e^{i\hat{q}_C\hat{p}_B},
\end{equation}
where $\hat{\cal{P}}_{CA}$ is the parity-swap operator\footnote{There is no physical meaning associated with the parity-swap operator. Instead it should be regarded as a mathematical tool for the consistency of the formulation.}, which when acting on momentum eigenstate of $C$ yields
\begin{equation}
    \label{swapOp}
    \hat{\cal{P}}_{CA}|p\rangle_C = |-p\rangle_A.
\end{equation}
Thus, the theory developed in Ref~\cite{Hoehn2018} is equivalent to that in Ref~\cite{Brukner}.

This concludes the overview of the first principle approach to switch QRFs via the perspective-neutral framework. The goal of this paper is to extend this theory to the quantum measurement formulation. We wish to develop the theory for quantum measurement starting from Dirac quantization, and follow the same symmetry reduction procedure to derive the reduced theory for a specific reference frame. It is also expected that by swapping between two QRFs, the measurement theory consistently recovers the results in the operational approach.

\section{Quantum Measurement}
\label{measurement}
To investigate how the quantum measurement process be embedded in a perspective-neutral framework, we employ the von Neumann quantum measurement theory, where a quantum measurement \replaced{is implemented by}{stages are separated into} two sub processes, Process 1 and Process 2.\deleted{as mentioned in the introduction section.} It is pointed out~\cite{Yang2018} that Process 1 must be described explicitly as observer-dependent\footnote{This statement plays a crucial role in the resolution of EPR paradox, as shown in ~\cite{Yang2018}.}. Different observers may have different descriptions of the same measurement event, which is vividly manifested by the Wigner's Friend thought experiment~\cite{Wigner, Wigner2}. An observer is associated with a specific QRF, this means Process 1 must be described specifically relative to a QRF. This is consistent with the fact that Process 1 is operational. An operational process should not be described in the perspective-neutral framework as the perspective-neutral framework does not admit an immediate operational interpretation~\cite{Hoehn2018}. Therefore, Process 1 needs to be described in the reduced Hilbert space, e.g., ${\cal{H}}_{AB|C}$ if particle C is the chosen QRF. On the other hand, \added{in this section we will show that the unitary} Process 2 can be \replaced{constructed through}{described in} the perspective-neutral framework, i.e., \replaced{from}{in} the Hilbert space ${\cal{H}}^{phys}$. 

With these guidelines in mind, our strategy is to formulate Process 2 in ${\cal{H}}^{phys}$, perform the redundancy reduction (or, symmetry reduction) procedure, and finally apply the projection operator representing Process 1 in the reduced Hilbert space.

\subsection{Measurement Theory for the Toy Model}
\label{gM}
The same toy model is used to develop the measurement theory. Besides particle A, B and C, another particle E is added as a measuring (or, auxiliary) particle. In this setup, we assume there is a pointer variable for particle E that is used to measure particles A and B. \added{Recall that a pointer variable is an observable of the measuring apparatus used to distinguish the outcomes at the end of the measurement process.} For simplicity, the momentum of particle E is chosen as pointer variable. We will derive the measurement formulation by initially choosing particle C as the QRF, then examine the theory by switching QRF from particle C to particle A.

The momentum constraint (\ref{constraint3}) is extended to include particle E, so that
\begin{equation}
\label{constraint4}
    \hat{P}|\psi\rangle^{phys} = (\hat{p}_A + \hat{p}_B + \hat{p}_C +\hat{p}_E)|\psi\rangle^{phys} = 0.
\end{equation}
\added{Supposed a complete set of measurement operators \{$\hat{\Lambda}_{p_E}\}$ defined in ${\cal{H}}^{kin}$ for particle A, B and C are measured with initial state $|\varphi\rangle^{kin}_{ABC}$, the essential question here is that whether we can construct a unitary operator to achieve Process 2.} The measurement formulation typically starts with the assumption that the initial state is a product state between the measurement apparatus system and the rest of the system. In ${\cal{H}}^{kin}$, the initial state is written as
\begin{equation}
    \label{initstate}
    |\psi\rangle^{kin} = |\varphi\rangle^{kin}_{ABC}|\phi_0\rangle_E,
\end{equation}
where we assume E is in an initial state of $|\phi_0\rangle_E$. The Process 2, where particle E interacts with the rest of system and becomes entangled with the rest of system, is described as unitary process in the von Neumann measurement theory. Denote the unitary operator in ${\cal{H}}^{kin}$ \added{to describe Process 2} as $\hat{U}$. \replaced{$\hat{U}$ is constructed by the following map}{$= e^{-i\hat{H}t}$, where $\hat{H}$ is the total Hamiltonian in ${\cal{H}}^{kin}$ that includes an interaction term between E and the measured particles. With these notations, Process 2 is formulated as}\footnote{Here we use the momentum eigenstate of particle E as pointer state for illustration purpose. It is not operational since such pointer state is maximally delocalized. A more practical scenario is \added{to} use other internal degree of freedom as pointer variable, which is discussed in Section \ref{AasQRF}.}
\begin{equation}
\label{Process2}
\begin{split}
   \hat{U}|\psi\rangle^{kin}  &= \hat{U}|\varphi\rangle^{kin}_{ABC}|\phi_0\rangle_E \\
   & = \int dp_E \hat{\Lambda}_{p_E}|\varphi\rangle^{kin}_{ABC}|p_E\rangle_E,
\end{split}
\end{equation}
\replaced{which implies that}{where} $\hat{\Lambda}_{p_E} = {}_E\langle p_E|\hat{U}|\phi_0\rangle_E$. \replaced{The proof that $\hat{U}$ defined by such map is a unitary operator can be found in Ref.~\cite{Nelsen}.}{is an operator defined in ${\cal{H}}^{kin}$ and only acts on particle A, B, and C.} Note that there are infinite number of $\hat{\Lambda}_{p_E}$ and they are labeled with index $p_E$. \added{We need to examine how this map is transformed when it is described in ${\cal{H}}^{phy}$ and the reduced Hilbert space, and whether it can lead to the correct measurement formulation in the reduced Hilbert space. First,} we apply the constraint map $\delta(\hat{P})$ to map the Hilbert space from ${\cal{H}}^{kin}$ to ${\cal{H}}^{phy}$ in both sides of (\ref{Process2})\footnote{Note this step of applying the constraint map $\delta(\hat{P})$ and the step of applying $\hat{U}$ in (\ref{Process2}) can be swapped, since the two operators are commutative.}, 
\begin{equation}
\label{Proc29}
\begin{split}
   |\psi\rangle^{phys}&:=\delta(\hat{P})\hat{U}|\psi\rangle^{kin}\\
   &= \delta(\hat{P}) \int dp_E \hat{\Lambda}_{p_E}|\varphi\rangle^{kin}_{ABC}|p_E\rangle_E.
\end{split}
\end{equation}
Now we start the symmetry reduction procedure by following the three steps described in previous section. First, particle C is picked as the reference frame. Next, a transformation operator, defined below,
\begin{equation}
    \label{transform2}
    \hat{T}_C = e^{i\hat{q}_C(\hat{p}_A+\hat{p}_B+\hat{p}_E)},
\end{equation}
is applied to (\ref{Process2}) after applying the constraint map,
\begin{equation}
\label{Proc22}
\begin{split}
    \hat{T}_C|\psi\rangle^{phys}= 
    \hat{T}_C\delta(\hat{P})\int dp_E \hat{\Lambda}_{p_E}|\varphi\rangle_{ABC}^{kin}|p_E\rangle_E.
\end{split}
\end{equation}
In Appendix \ref{AppendixA}, we show that
\begin{equation}
\label{Proc23}
\begin{split}
   &|\psi\rangle_{C,ABE} := \hat{T}_C|\psi\rangle^{phys}  = |p=0\rangle_C\otimes\\
   &\int dp_E \int dp_Adp_B\chi_{p_E}(p_A, p_B)|p_A\rangle|p_B\rangle|p_E\rangle_E.
\end{split}
\end{equation}
where $\chi_{p_E}(p_A, p_B)$ is defined in (\ref{A4}). If we further assume that $|\psi\rangle_{ABC}^{kin}=|\varphi\rangle_{AB}|\xi\rangle_C$ is a product state, it is shown in Appendix \ref{AppendixA} that in this case,
\begin{equation}
    \label{Proc24}
    |\psi\rangle_{C,ABE}=|p=0\rangle_C\otimes \int dp_E \hat{\Gamma}_{p_E} |\varphi\rangle_{AB}|p_E\rangle_E.
\end{equation}
where operator $\hat{\Gamma}_{p_E}$ only acts on particle A and B and is defined in (\ref{Gamma2}). As the final step of the reduction procedure, we discard the redundant state for particle C.
\begin{equation}
    \label{state5}
    \begin{split}
    |\psi\rangle_{ABE|C} &:=\sqrt{2\pi}\langle q_C=0|\psi\rangle_{C,ABE} \\
    & = \int dp_E \hat{\Gamma}_{p_E} |\varphi\rangle_{AB|C}|p_E\rangle_E.
    \end{split}
\end{equation}
At this point, the pointer variable $p_E$ is well-defined in the reduced Hilbert space ${\cal{H}}_{ABE|C}$. The Process 1 in von Neumann measurement theory is described by applying the projection
\begin{equation}
    \label{Projector}
    \hat{P}_m = |p_m\rangle_E\langle p_m|
\end{equation}
to (\ref{state5}), resulting
\begin{equation}
\label{Process1}
    |\psi\rangle^m_{ABE|C} := \hat{P}_m|\psi\rangle_{ABE|C} = \hat{\Gamma}_{p_m}|\varphi\rangle_{AB|C}|p_m\rangle_E.
\end{equation}
Dropping the pointer state for particle E, and defining $\hat{M}_m = \hat{\Gamma}_{p_m}/\sqrt{N}$, where $N$ is a normalization factor, we further rewrite the state vector for particle A and B after measurement as the well-known form,
\begin{equation}
    \label{state3}
    |\varphi\rangle_{AB|C}^m = \frac{ \hat{M}_m|\varphi\rangle_{AB|C}}{\sqrt{\langle\varphi|\hat{M}_m^\dag\hat{M}_m|\varphi\rangle_{AB|C}}}.
\end{equation}
The probability of the measurement outcome for $m$ is
\begin{equation}
    \label{probC}
    \rho_{m|C} = {}_{AB|C}\langle\varphi|\hat{M}_m^\dag\hat{M}_m|\varphi\rangle_{AB|C}.
\end{equation}
Note that index $m$ refers to $p_m$ which in our toy model is a continuous real number $p_m \in \cal{R}$ for particle E. Appendix \ref{AppendixC} verifies that the set of operators $\{\hat{M}_m\}$ satisfy the completeness condition, i.e.,
\begin{equation}
    \label{completeness}
    \int dp_m \hat{M}_{p_m}^\dag\hat{M}_{p_m} = \hat{I}_{AB|C},
\end{equation}
where $\hat{I}_{AB|C}$ is a unit operator in Hilbert space ${\cal{H}}_{AB|C}$.

We can also write down the relation between operators $\hat{M}_{p_m}$ and $\hat{\Lambda}_{p_m}$ as
\begin{equation}
   \label{opM}
   \begin{split}
    \hat{M}_{p_m} & = \sqrt{\frac{2\pi}{N}}\langle q_C=0|e^{i\hat{q}_C(\hat{p}_A+\hat{p}_B+p_m)} \\
    & \times\delta(\hat{p}_A+\hat{p}_B+\hat{p}_B+p_m)\hat{\Lambda}_{p_m}|\xi\rangle_C.
    \end{split}
\end{equation}
The derivations process in this Section can be understood reversely. Suppose one wants to measure a Hermitian observable $\hat{O}_{AB}$ in reduced Hilbert space ${\cal{H}}_{ABE|C}$ using $\hat{p}_E$ as pointer variable. $\hat{O}_{AB}$ can be first decomposed to a complete set of operators $\{\hat{M}_{p_m}\}$ through the eigenvalue decomposition such that $\int dp_m\hat{M}_{p_m}^\dag\hat{M}_{p_m} = \hat{I}_{AB|C}$. From (\ref{opM}), one can reversely derive a set of operators \{$\hat{\Lambda}_{p_m}\}$ in ${\cal{H}}^{kin}$, and then construct the unitary operator through the map defined in (\ref{Process2}). Thus, we show that the measurement process for observable $\hat{O}_{AB}$ in ${\cal{H}}_{ABE|C}$ can be implemented by a projection operation in ${\cal{H}}_{ABE|C}$, and a unitary process embedded in the perspective-neutral framework. An important consequence is that it allows us to derive the measurement formulation when switching to a different quantum reference frame.

\subsection{Switching QRFs}
\label{switchQRFM}
Now we switch the QRF to be particle A and examine how the measurement formulation changes. Following the same derivation procedure \added{but not assuming a product state between system AB and C as in deriving (\ref{state3})}, we find the final state for particle B and C after measurement is
\begin{equation}
    \label{state6}
    \begin{split}
    |\varphi\rangle_{BC|A}^m & = \frac{1}{\sqrt{\rho_{m|A}}}\int dp_Bdp_C\kappa_{p_m}(p_B, p_C)|p_B\rangle|p_C\rangle \\
    & = \frac{1}{\sqrt{\rho_{m|A}}}|\kappa\rangle_{BC|A}^{p_m},
    \end{split}
\end{equation}
where 
\begin{equation}
    \begin{split}
     &\kappa_{p_m}(p_B,p_C) =\int dp'_Adp'_Bdp'_C \lambda_{p_m} \psi_{ABC}(p'_A, p'_B, p'_C) \\
     &\lambda_{p_m} = \int u(-p_B-p_C-p_m, p_B, p_C,p_m; p'_A, p'_B, p'_C, p'_E) \\
    &\times \phi_0(p'_E)dp'_E \\ 
    &\rho_{m|A} = {}_{BC|A}\langle\kappa|\kappa\rangle_{BC|A}^{p_m}.
    \end{split}
\end{equation}
and matrix element $u$ is defined in (\ref{A1}). From (\ref{state3}) and (\ref{state6}), we can derive how the two state vector $|\varphi\rangle_{BC|A}^m$ and $|\varphi\rangle_{AB|C}^m$ are connected. Define a transformation operator
\begin{equation}
    \label{swapOp2}
    \hat{\cal{S}}=\hat{\cal{P}}_{AC}e^{i\hat{q}_A(\hat{p}_B+\hat{p}_E)},
\end{equation}
where $\hat{\cal{P}}_{AC}$ is the parity swap operator defined in (\ref{swapOp}). In Appendix \ref{AppendixB}, we prove that
\begin{equation}
    \label{swapSate2}
    |\varphi\rangle_{BC|A}^m = \hat{\cal{S}}_m|\varphi\rangle_{AB|C}^m,
\end{equation}
where $\hat{\cal{S}}_m=\hat{\cal{P}}_{AC}e^{i\hat{q}_A(\hat{p}_B+p_m)}$. 

As also shown in Appendix \ref{AppendixB}, the probability of measurement outcome $m$ from the perspective of QRF A is conserved, i.e., $\rho_{m|A}=\rho_{m|C}$. The conservation of the probabilities of the same measurement outcome from perspectives of both QRFs is a natural consequence of the formulation, owning to the unitary property of operator $\hat{\cal{S}}_m$. This result is consistent with that in Ref~\cite{Brukner}.

It it important to note that the derivation of (\ref{state6}) depends on two assumptions. First, the observer associated with QRF A knows the same measurement outcome $m$ as the observer associated with QRF C. This should not be taken for granted if A and C are separated remotely. The synchronization of the information regarding the measurement outcome is necessary~\cite{Yang2018}. Second, the pointer variable projection $\hat{P}_m$, defined in (\ref{Projector}), is invariant under the \replaced{switching}{transformation} of QRF. One can easily verify that
\begin{equation}
    \label{invariant}
    \hat{\cal{S}}\hat{P}_m\hat{\cal{S}}^\dag = \hat{P}_m.
\end{equation}

To demonstrate the effect of $\hat{\cal{S}}_m$, we consider a concrete example. Supposed the measurement operator $\hat{M}_m$ projects the state for particle A and B into a product state from C perspective, i.e.,
\begin{equation}
    \hat{M}_m = |\chi_m\rangle_A|\varphi_m\rangle_B{\otimes}_A\langle\chi_m|{}_B\langle\varphi_m|.
\end{equation}
The final state after measurement, according to (\ref{state3}), is
\begin{equation}
    |\phi\rangle_{AB|C}^m = |\chi_m\rangle_A|\varphi_m\rangle_B.
\end{equation}
Thus, from C perspective, A, B, and E are not entangled with each other after measurement. Now if we switch the QRF from C to A, from A perspective, the observed probability and the value of pointer variable corresponding to $m$ are the same, but the quantum state for B and C after measurement is
\begin{equation}
    \label{state7}
    \begin{split}
    |\phi\rangle_{BC|A}^m &= \hat{\cal{S}}_m |\chi_m\rangle_A|\varphi_m\rangle_B \\
    &=\hat{\cal{S}}_m\int dp_Adp_B \chi_m(p_A)\varphi_m(p_B)|p_A\rangle|p_B\rangle\\
    &= \int dp_Bdp_C \chi_m(-p_B-p_C-p_m) \\
    & \times\varphi_m(p_B)|p_B\rangle|p_C\rangle.
    \end{split}
\end{equation}
It shows that B and C are entangled.  This result is similar to that in Ref~\cite{Brukner}, except that the auxiliary system E is not necessarily entangled with B or C.

\subsection{Measurement Apparatus as QRF}
\label{AasQRF}
So far the formulation assumes the measuring system (or, auxiliary system) and the QRF are two different systems. However, there is an important situation where we need to treat the measuring system itself as the QRF. For instance, supposed system A is the laboratory, systems B and C are moving with a same speed relative to A, and C is measuring B. It is more difficult to describe the measurement process from A perspective, particularly when the pointer variable depends on the speed. It is much easier to \replaced{describe}{description} the measurement process first at the rest reference frame and then transform to describe the process from A perspective. In this case, it means we take the measuring system C also as the QRF, derive the measurement formulation from C perspective, then transform back to the perspective of A as QRF\footnote{This technique has been proposed to derive the operation meaning of a relativistic spin system~\cite{BruknerSpin}.}. In Ref.~\cite{Brukner}, such scenario is considered where system C has both external degree of freedom (i.e., the momentum degree of freedom) that is discarded when A is taken as a QRF, and the internal degree of freedom that is acting as a pointer variable to measure system B.

In the toy model considered in this paper, particles A, B, and C are interacting and only constrained with a global translational invariance. Relative speed is not considered in this simple model. We assume particle C has both external degree of freedom, and internal degree of freedom that acts as a pointer variable. Thus, a general state for particle C, before considering the constraint, is
\begin{equation}
    \label{stateC}
    |\phi\rangle_C^{kin} = \sum_m \int dp_C \phi_m(p_C)|p_C, \sigma_m\rangle_C.
\end{equation}
An eigenstate associated with a particular value of pointer variable $\sigma_m=\sigma_n$ is
\begin{equation}
    \label{stateC2}
    |\phi_n\rangle_C = \int dp_C \phi_n(p_C)|p_C, \sigma_n\rangle_C.
\end{equation}
The state vector satisfies the following orthogonal identities:
\begin{equation}
    \label{Ortho}
    \begin{split}
      \langle p'_C, \sigma_n |p_C, \sigma_m \rangle &= \delta_{mn}\delta(p'_C - p_C),\\
      \langle \phi_n | \phi_m\rangle &= \delta_{mn}.
    \end{split}
\end{equation}
The pointer variable (i.e., the internal degree of freedom), which can be the energy level, or the spin, in theory can depend on the momentum variable. A good example is the spin of a relativistic particle \added{that} can depend on the velocity of the particle. For simplicity we assume here that the pointer variable is independent of the momentum variable. 

Similar to (\ref{initstate}), we assume the initial state of A, B, and C is a product state in ${\cal{H}}^{kin}$, written as
\begin{equation}
    \label{initstate2}
    |\psi\rangle^{kin} = |\varphi\rangle_{AB}|\phi_0\rangle_C.
\end{equation}
The Process 2, where particle C interacts with B, is described as a unitary process. The unitary operator $\hat{U}^{kin}$ is associated with the Hamiltonian $\hat{H}^{kin}$ defined in ${\cal{H}}^{kin}$.
\begin{equation}
\label{Proc25}
\begin{split}
   \hat{U}^{kin}|\psi\rangle^{kin}  &= \hat{U}^{kin}|\varphi\rangle_{AB}|\phi_0\rangle_C \\
   & = \sum_m \hat{\Gamma}_{m}^{kin}|\varphi\rangle_{AB}|\phi_m\rangle_C,
\end{split}
\end{equation}
where $\hat{\Gamma}_m = {}_C\langle \phi_m|\hat{U}|\phi_0\rangle_C$ is an operator defined in ${\cal{H}}^{kin}$ and only acts on particle A and B. Next we apply the constraint map $\delta(\hat{P})$ to (\ref{Proc25}) and get the physical state in Hilbert space ${\cal{H}}^{phys}$, (\ref{Proc25}) is changed to
\begin{equation}
\label{Proc26}
\begin{split}
   |\psi\rangle^{phys}&=\delta(\hat{P})\hat{U}|\psi\rangle^{kin}  \\
   &= \delta(\hat{P}) \sum_m \hat{\Gamma}_{m} |\varphi\rangle_{AB}|\phi_m\rangle_C.
\end{split}
\end{equation}
To proceed with the redundancy reduction procedure, particle C is picked as the reference frame. The transformation operator $\hat{T}_C$ is simply defined below\footnote{Operator $\hat{T}_C$ is independent of the pointer variable $\sigma$ due to the assumption that the pointer variable is invariant \replaced{when switching QRFs}{under translation}. In a more complicated model, such assumption may not hold. See more discussion in Section \ref{Orders}}
\begin{equation}
    \label{transform4}
    \hat{T}_C = e^{i\hat{q}_C(\hat{p}_A+\hat{p}_B)}.
\end{equation}
In Appendix \ref{AppendixD}, we show that after applying $\hat{T}_C$ to (\ref{Proc26}), we obtain
\begin{equation}
\label{Proc27}
\begin{split}
    |\psi\rangle_{C,AB} & =\hat{T}_C|\psi\rangle^{phys} \\
    &= \sum_m|p=0, \sigma_m\rangle_C \otimes \hat{M}_{m}|\varphi\rangle_{AB},
\end{split}
\end{equation}
where operator $\hat{M}_m$ is defined in (\ref{D5}). The last step of removing the redundancy of external degree of freedom of QRF C can be combined with the projection of the Process 1 of the measurement. Namely, for a measurement outcome corresponding to pointer variable of $\sigma_n$, we can apply $\sqrt{2\pi}\langle q_C=0, \sigma_n|\otimes$ to (\ref{Proc27}), and obtain the final state after the measurement
\begin{equation}
    \label{Proc15}
    \begin{split}
    |\varphi\rangle_{AB|C}^n &= \sqrt{2\pi}\langle q_C=0, \sigma_n|\psi\rangle_{C,AB}\\
    &= \hat{M}_{n}|\varphi\rangle_{AB|C}.
    \end{split}
\end{equation}
Normalization of the above state vector gives
\begin{equation}
    \label{state7}
    |\tilde{\varphi}\rangle_{AB|C}^n = \frac{ \hat{M}_n|\varphi\rangle_{AB|C}}{\sqrt{\langle\varphi|\hat{M}_n^\dag\hat{M}_n|\varphi\rangle_{AB|C}}}.
\end{equation}
\added{Thus, we show that in the case of choosing the measurement apparatus as QRF, the measurement process  in the reduced Hilbert space ${\cal{H}}_{ABE|C}$ can be implemented by a projection operation in ${\cal{H}}_{ABE|C}$, and a unitary process embedded in the perspective-neutral framework.} Next we want to know how the same measurement process is perceived if we choose particle A as QRF. In Appendix \ref{AppendixE}, we show that starting from the same decomposition (\ref{Proc25}), after carrying over the redundancy reduction procedure with particle A as the QRF, and with transformation operator defined as $\hat{T}_A = e^{i\hat{q}_A(\hat{p}_B+\hat{p}_C)}$, we get
\begin{equation}
\label{Proc28}
\begin{split}
   |\varphi\rangle_{BC|A} & = \sum_m \int dp_Bdp_C\phi_m(p_C)\chi_m(p_B, p_C) \\
   & \times|p_B\rangle|p_C, \sigma_m\rangle,
\end{split}
\end{equation}
where $\chi_m(p_B, p_C)$ is defined in (\ref{E3}). The question here is that what \deleted{the} projection operator for Process 1 we should apply at this point. Since we assume the pointer variable does not transform under translation, from A perspective, the pointer variable is the same as from C perspective. Thus, we define the projection operator from C perspective as $\hat{P}_n = |\sigma_n\rangle_C\langle\sigma_n|$, assuming that the pointer variable $\sigma_m$ is separable from the momentum degree of freedom $p_C$, i.e., $|p_C, \sigma_m\rangle=|p_C\rangle| \sigma_m\rangle$. Applying $\hat{P}_n$ to (\ref{Proc28}), we obtain the final state after measurement,
\begin{equation}
    \label{Proc16}
    \begin{split}
    &|\varphi\rangle_{BC|A}^n = \hat{P}_n|\varphi\rangle_{BC|A}\\
    &= \int dp_Bdp_C\phi_n(p_C)\chi_n(p_B, p_C)|p_B\rangle|p_C\rangle|\sigma_n\rangle.
    \end{split}
\end{equation}
(\ref{Proc16}) shows that in the final state from A perspective, the degree of freedom of particle B is entangled with external degree of freedom of C, but disentangled with the internal degree of freedom of C, i.e., the pointer variable of C. This reproduces the same conclusion in Ref~\cite{Brukner}. Defining transformation map
\begin{equation}
    \label{swapOp3}
    \begin{split}
    \hat{\cal{S}}_n &= \hat{\cal{P}}^n_{AC}e^{i\hat{q}_A\hat{p}_B} \quad \text{where}\\
    \hat{\cal{P}}^n_{AC}|p\rangle_A &= |-p, \sigma_n\rangle_C.
    \end{split}
\end{equation}
We show in Appendix \ref{AppendixF} that (\ref{Proc16}) and (\ref{Proc15}) can be transformed similar to (\ref{swapSate2}) 
\begin{equation}
    \label{swapSate3}
    |\varphi\rangle_{BC|A}^n = \hat{\cal{S}}_n|\varphi\rangle_{AB|C}^n,
\end{equation}
It follows immediately that the probability of measurement outcome correspondent to pointer variable value $\sigma_n$ is \deleted{the} preserved from the perspective of either QRF.

\section{Discussion and Conclusion}
\label{discussion}

\subsection{The Order of Reduction and Projection}
\label{Orders}
The projection in the  Process 1 of the von Neumann quantum measurement theory is an operational process, i.e., a process that can be confirmed or observed through the pointer variable of the measurement apparatus. The value of the pointer variable is read and recorded with respect to a specific QRF. Thus, it should be described in the reduced Hilbert space with respect to that particular QRF. This means we perform the symmetry reduction of the Dirac quantization first, then the projection operation. As a consequence, the operational meaning of the pointer variable is well-defined in the reduced Hilbert space. Such a procedure is consistent with the expectation that a measurement event should be operationally well-defined. An opposite procedure is to perform the projection first in the perspective-neutral Hilbert space, then the symmetry reduction procedure. However, the projection operator can be transformed during the reduction procedure, and the operational meaning of that pointer variable becomes difficult to define. For instance, let's consider particle A, B, C are relativistic particles with spins and moving with different speeds. Supposed the spin of particle C is the pointer variable for a measurement on particle B and C itself is also the QRF. The operation meaning of spin is well defined in a rest QRF with particle C. On the other hand, it is a challenge to perform the projection operation in the perspective-neutral formulation of the physical Hilbert space. This is because the spin of particle C depends on its momentum and particle C can be in a superposition state of momentum. The symmetry reduction procedure involves transformation of the momentum operator which in turn transforms the spin. How to define a projection operator in the perspective-neutral Hilbert space that can be transformed to the reduced Hilbert space and yield the well-defined spin projector is not yet considered and needs further investigation\footnote{Ref~\cite{BruknerSpin} provides a solution for defining the operational meaning of the spin of a relativistic particle when swapping QRFs that are moving with a speed relative to each other. However, Ref~\cite{BruknerSpin} is not based on the perspective-neutral framework and of course does not address the transformation from the perspective-neutral Hilbert space to the reduced Hilbert space.}.

There is, however, a special situation when the pointer variable is invariant in the reduction procedure. In this case, the order of reduction and projection does not impact the formulation of measurement. Considered the example in Section \ref{AasQRF}, where we perform reduction first, and projection later to derive the measurement formulation. But in fact the order can be reserved and yield the same results. Supposed we first apply the projection operator $\hat{P}'_n = |\phi_n\rangle_C\langle\phi_n|$ to (\ref{Proc25}), and obtain
\begin{equation}
\label{Proc29}
   \hat{P}'_n\hat{U}^{kin}|\psi\rangle^{kin}  
   = \hat{\Gamma}_{n}^{kin}|\varphi\rangle_{AB}|\phi_n\rangle_C.
\end{equation}
Then we proceed the reduction procedure by choosing C as QRF, and obtain (\ref{Proc15}). If choosing A as QRF, the reduction process resulting with 
\begin{equation}
    \label{Proc19}
    \begin{split}
    |\varphi\rangle_{BC|A}^n &= \int dp_Bdp_C\phi_n(p_C)\chi_n(p_B, p_C) \\
    &\times|p_B\rangle|p_C,\sigma_n\rangle,
    \end{split}
\end{equation}
which is essentially the same as (\ref{Proc16}). The reason for this is that the pointer variable, $\sigma_m$, is invariant in the symmetry reduction procedure, and consequently also invariant when switching QRFs. However, this is a special case and should not be considered true in general.

\subsection{Synchronization Among Different QRFs}
When deriving (\ref{state6}) or (\ref{Proc16}), we assume that the observer associated with QRF A knows the measurement outcome that is inferred from the pointer variable $\sigma_n$ associated with QRF C. In other words, we assume that the measurement projection operator $\hat{P}_n$ is known to both observers. This assumption should not be taken for granted. Quantum measurement must be described relative to the local observer. An observer who does not access to the measurement results will not have the complete information and can only describe the system up to the level of previous knowledge that the observer has. To ensure the descriptions of different observers are consistent, different observers should synchronize information regarding the measurement results in order to have consistent descriptions of a quantum system~\cite{Yang2018, Yang2019}. The results in this paper shows that a further refinement on the information synchronization is needed. Specifically, the pointer variable needs to be transformed properly when switching QRFs. In the special case that the pointer variable, consequently the projection operator $\hat{P}_n$, are invariant from either A perspective or C perspective, we simply use the same operator $\hat{P}_n$ for the projecting operation. This is what we have done when deriving (\ref{state6}) or (\ref{swapSate3}). 

To illustrate this point more concretely, we go back to the example in Section \ref{AasQRF} where particle C is the measurement apparatus and we choose particle A as QRF. After the symmetry reduction procedure we arrive at (\ref{Proc28}) and need to apply a projection operator for particle C from A perspective. Now, instead of choosing $\hat{P}_{n|A}=|\sigma_n\rangle_C\langle\sigma_c|$, one may attempt to assign the projection operator $\hat{P}'_{n|A} = |\phi_n\rangle_C\langle\phi_n|$. \replaced{One possible reason to make this choice}{One reason this choice seems make sense} is that an arbitrary wave function for particle C that is associated with pointer variable $\sigma_n$ is given by (\ref{stateC2}). Applying this projection operator to (\ref{Proc28}), we obtain a final state after measurement,
\begin{equation}
    \label{Proc17}
    \begin{split}
    |\varphi'\rangle_{BC|A}^n &= \hat{P}'_{n|A}|\varphi\rangle_{BC|A}\\
    &= (\int dp_B\chi'_n(p_B)|p_B\rangle)|\phi_n\rangle_C,\\
    \chi'_n(p_B)&=\int dp_C|\phi_n(p_C)|^2\chi_n(p_B, p_C).
    \end{split}
\end{equation}
Obviously, $|\varphi'\rangle_{BC|A}^n$ is a product state. Particle B is not entangled with either the external degree of freedom or the internal pointer variable of particle C. This result is different from (\ref{Proc16}) where particle B is entangled with the external degree of freedom of C. Further calculation shows that the probability of the measurement resulting from $\hat{P}'_{n|A}$ is not the same as the probability from C perspective. The preservation of the measurement probability from different QRFs appears to be a natural criteria that should be satisfied. Thus, the choice of projection operator $\hat{P}'_{n|A}$ is incorrect.

This observation can be further explained as following. If from C perspective the projection operator is $\hat{P}_{n|C}=|\sigma_n\rangle_C\langle\sigma_c|$, then from A perspective, the corresponding projection operator must be derived via proper transformation, $\hat{P}_{n|A}=\hat{\cal{S}}_n(\hat{P}_{n|C})\hat{\cal{S}}_n^\dag$ where $\hat{\cal{S}}_n$ is defined in (\ref{swapOp3}). Since $\hat{\cal{S}}_n$ has no effect on $|\sigma_n\rangle_C\langle\sigma_c|$, we obtain $\hat{P}_{n|A}=\hat{P}_{n|C}$. This rules out $\hat{P}'_{n|A}$ as the the correct projection operator.

We can summarize the implications of the relational formulation of quantum measurement based on Refs~\cite{Yang2018, Yang2019} and this paper as following.
\begin{enumerate}
    \item \textit{Measured reality is relative.} Information obtained through quantum measurement is local. Measurement must be described explicitly relative to the local observer.  
    \item \textit{Synchronization of local reality.} Different observers should synchronize information regarding the measurement results in order to have consistent descriptions of a quantum system.
    \item When the measurement information \replaced{is}{being} synchronized among different observers, proper transformation on the projection operator must be performed to ensure consistency.
\end{enumerate}

Applying the first implication, we are able to resolve the EPR paradox~\cite{EPR, Yang2018}. In that resolution, a quantum measurement should be explicitly described as observer dependent. The idea of observer-independent quantum state is abandoned since it depends on the assumption of Super Observer. By recognizing that the element of physical reality obtained from local measurement is only valid relatively to the local observer, the completeness of quantum mechanics and locality can coexist~\cite{Yang2018}. Applying the second implication, we are able to resolve the Wigner's friend paradox and the extended version~\cite{FR2018, Yang2019}. These thought experiments provide clear example for the need of information synchronization in order to achieve a consistent description of a quantum system by different observers. Ref~\cite{Brukner2} shows similar idea that the assumption of observer independent fact cannot resolve the Wigner's friend type of paradox. Latest experiment appears to confirm that observer-independent description of a quantum system must be rejected~\cite{Proietti19}. Lastly, the third implication is a \replaced{confirmation}{refinement} of the second implication, and is just discussed in this section.

\subsection{The Wigner-Araki-Yanase Theorem}
\added{
One of the fundamental questions in quantum measurement theory is whether a quantum measurement procedure can be implemented for a given Hermitian operator. When there is an additive conservation law, the Wigner-Araki-Yanase (WAY) Theorem~\cite{Wigner3, Araki, Ahmadi1} imposes restriction on the measurement procedure in that only observables that are commutative with the conserved quantity can be implemented with a repeatable measurement. More precisely, suppose a quantum system S is measured with an apparatus A and the correspondent Process 2 is described by a unitary operator $\hat{U}$. Suppose also there is an additive conservation quantity, denoted as $\hat{L}=\hat{L}_S\otimes\hat{I}_A+\hat{I}_S\otimes\hat{L}_A$ where $\hat{I}$ is a unit operator, such that $[\hat{U}, \hat{L}]=0$. Then, the WAY theorem states that the only observables $\hat{O}_S$ for which it is possible to implement the projective measurement are those that are commutative with $\hat{L}_S$, i.e. those that satisfy $[\hat{O}_S, \hat{L}_S]=0$. 
}

\added{
One may ask whether the WAY theorem imposes restriction on the derivation of measurement formulation in this paper. We can examine the derivation in Section \ref{gM}. The conservative quantity in ${\cal{H}}^{phys}$ is the total momentum $\hat{P}=(\hat{p}_A+\hat{p}_B+\hat{p}_C)+\hat{p}_E$, which is the constraint and satisfies $[\hat{P}, \hat{U}^{phys}]=0$. We can consider $\hat{L}_S=\hat{p}_A+\hat{p}_B+\hat{p}_C$ and $\hat{L}_A=\hat{p}_E$. However, ${\cal{H}}^{phys}$ is a perspective neutral Hilbert space and as discussed earlier, one cannot consider the projective Process 1 in this Hilbert space. In other words, it is not required that the pointer variable states are orthogonal in ${\cal{H}}^{phys}$. Conceptually we cannot implement a complete von Neumann measurement procedure including both Process 1 and 2 in ${\cal{H}}^{phys}$. It is conceptually incorrect to apply the WAY theory in ${\cal{H}}^{phys}$. On the other hand, in the reduced Hilbert space, such as in ${\cal{H}}_{ABE|C}$, the conservative quantity does not exist because once taken particle C as the QRF, the translation symmetry is broken. Thus, the WAY theory cannot be applicable in any reduced Hilbert space either. We conclude that the WAY theorem is not applicable to the formulation presented here.
}

\added{
However, the spirit behind the WAY theorem, i.e., given a Hermitian operator and a constraint, whether a unitary operator can be constructed to implement Process 2 must be carefully examined. In the context of the toy model, this means that given a Hermitian operator in the reduced Hilbert space ${\cal{H}}_{ABE|C}$, we need to confirm that there exists a unitary operator in ${\cal{H}}^{kin}$ that can implement Process 2. The last paragraph in Section \ref{gM} shows this claim is indeed correct.
}
\subsection{Limitation}
The toy model used in this paper is relatively \replaced{simple}{simply}. The systems in this model are one dimensional and only have constraint due to the translational symmetry. The simple model allows one to construct a globally valid gauge fixing condition. Ref.~\cite{Hoehn2018-2} extends the method to three dimensional N-body systems that have both translational and rotational symmetries. It will be interesting to confirm that the measurement theory developed in this paper will be applicable to the three dimensional N-body systems as well. However, as Ref.~\cite{Hoehn2018-2} points out, for three dimensional N-body systems that have both translational and rotational symmetries, one cannot find globally valid gauge fixing conditions. Nevertheless, we expect that the measurement formulation presented in this paper is applicable to three dimensional N-body systems based on the formulation in Ref.~\cite{Hoehn2018-2}. It is desirable to extend the theory to even more complicated model that includes other degree of freedom such as spin. 

The assumption that the pointer variable is invariant during the reduction process is another major limitation. \added{This is particularly true in the case that the measurement apparatus is also the QRF where pointer variable $\hat{\sigma}_m$ is invariant when switching QRF. If considered $\hat{\sigma}_m$ is spin of particle C, such assumption is reasonable in a non-relativistic case. However in a relativistic case, spin depends on momentum and} \deleted{If the pointer variable} changes when switching QRFs or during the reduction process, the transformation operator such as $\hat{T}_C$ needs to include the pointer variable, and the formulation will be more complicated. \deleted{We would also expect that such condition should manifest the importance of the order of reduction versus projection, as discussed in Section \ref{Orders}.} \added{In the present paper, since we assume the pointer variable is invariant when switching QRFs, the transformation of this pointer variable when switching QRFs is in a sense a priori known. Thus, the promised of not relying on a priori known transformation in the first principle approach is not completely fulfilled. Nevertheless, the transformation of other degree of freedoms such as momentum of the QRF can be derived. Once the dependency of pointer variable $\hat{\sigma}_m$ is factored into $\hat{T}_C$ during the reduction process, such limitation can be overcome.}

The projection process in the von Neumann quantum measurement theory is a simple mathematical modeling of the actual measuring process. It cannot explain the mechanism of ``wave function collapse''. Ref.~\cite{Hoehn2018} speculates that by including the measurement interaction into the perspective-neutral structure, it may possibly lead to the ``collapse'' in the respective internal perspective. Obviously our formulation here does not achieve such goal. However, we do bring in new conceptual implications discussed earlier.

\subsection{Conclusions}
Inspired by the novel approach of switching QRFs via a perspective-neutral framework~\cite{Hoehn2018}, this paper extends the approach to the quantum measurement process. Specifically, we show the von Neumann quantum measurement theory can be embedded in the perspective-neutral framework. Based on the same simple toy model as in Ref.~\cite{Hoehn2018}, we show that Process 2 in the von Neumann measurement theory, which is a unitary process, can be formulated in the perspective-neutral framework, while Process 1, which is a projection, should be described after the perspective-neutral structure is reduced to be specifically relative to a QRF. In the special case when the pointer variable is invariant to the redundancy reduction procedure (hence invariant when switching QRFs), the order of reduction versus projection has no impact on the results. Our results successfully reconstruct the measurement formulation, as shown in Section \ref{gM} and \ref{switchQRFM}, from perspectives of different QRFs. This allows us to further derive the transformation operator for the measurement outcome when switching QRFs, given in (\ref{swapOp2}). These results are consistent with that in Ref.~\cite{Brukner}, with the advantage of being derived from the first principles proposed in Ref.~\cite{Hoehn2018}. 

Furthermore, when the measurement apparatus itself is considered as a QRF, our measurement formulation provides additional conceptual implications. In particularly, when switching QRFs, the projection operator \deleted{in the} must be transformed properly, otherwise it causes inconsistency. For instance, the probabilities from different observers for the same measurement outcome can be different. Conceptually, this further \replaced{confirms}{refines} the synchronization principle~\cite{Yang2018, Yang2019} that is used to resolve paradox for the extended Wigner's friend thought experiment. 

In conclusion, the results presented in this paper further confirm the validity of the first principle approach of switching QRFs via a perspective-neutral framework~\cite{Hoehn2018}. Our formulations on the measurement theory fills in the gap to recover the transformation mechanism when switching QRFs using the operational approach~\cite{Brukner}. All these research results together extend the understanding on how quantum systems should be described when the reference frame itself is a quantum system and when switching QRFs.

\begin{acknowledgements}
I would like to acknowledge the contributions from Philipp A. H\"{o}hn through many insightful discussions that help to refine some of the ideas in the paper, and sincerely thank him for pointing out technical errors during the preparation of the manuscript. \added{I would also like to  thank the anonymous referees for the valuable comments, in particular, for bringing up the discussion of the Wagner-Araki-Yanase theorem.}
\end{acknowledgements}

\bibliographystyle{plain}

\onecolumn
\newpage
\appendix

\section{Proof of Eq.(\ref{Proc23})}
\label{AppendixA}
In general, the unitary operator $\hat{U}$ can be decomposed as
\begin{equation}
    \label{A1}
    \begin{split}
    \hat{U}& = \int dp_Adp_Bdp_Cd\tilde{p}_Edp'_Adp'_Bdp'_Cd\tilde{p}'_E
    u(p_A, p_B, p_C, \tilde{p}_E; p'_A, p'_B, p'_C, \tilde{p}'_E)\\
    & \times |p_A\rangle|p_B\rangle|p_C\rangle|\tilde{p}_E\rangle\langle p'_A|\langle p'_B|\langle p'_C|\langle \tilde{p}'_E|.
    \end{split}
\end{equation}
The decomposition of operator $\hat{\Lambda}_{p_E}$ reads
\begin{equation}
    \label{A2}
    \begin{split}
    \hat{\Lambda}_{p_E}&=\langle p_E|\hat{U}|\phi_0\rangle_E = \int dp_Adp_Bdp_Cdp'_Adp'_Bdp'_C\lambda_{p_E} |p_A\rangle|p_B\rangle|p_C\rangle_C\langle p'_A|\langle p'_B|\langle p'_C|, \\
    \lambda_{p_E} &= \int u(p_A, p_B, p_C, p_E; p'_A, p'_B, p'_C, p'_E)\phi_0(p'_E)dp'_E.
    \end{split}
\end{equation}
Insert (\ref{A2}) to the R.H.S of (\ref{Proc22}), we obtain
\begin{equation}
    \label{A3}
    \begin{split}
    |\psi\rangle_{C,ABE} &= \hat{T}_C\int dp_Adp_Bdp_Cdp'_Adp'_Bdp'_Cdp_E\delta(p_A+p_B+p_C+p_E)\lambda_{p_E}|p_A\rangle|p_B\rangle|p_C\rangle_C|p_E\rangle \\
    &\times \langle p'_A|\langle p'_B|\langle p'_C|\psi\rangle_{ABC}\\
    & = |p=0\rangle_C\otimes\int dp_E |p_E\rangle \int dp_Adp_Bdp'_Adp'_Bdp'_C\lambda_{p_E}|p_A\rangle|p_B\rangle\langle p'_A|\langle p'_B|\langle p'_C|\psi\rangle_{ABC}\\
    & = |p=0\rangle_C\otimes\int dp_Edp_Adp_B \chi_{p_E}(p_A,p_B)|p_A\rangle|p_B\rangle|p_E\rangle,
    \end{split}
\end{equation}
where $\chi_{p_E}(p_A,p_B)$ is defined as
\begin{equation}
    \label{A4}
    \begin{split}
    \chi_{p_E}(p_A,p_B)  &= \int dp'_Adp'_Bdp'_C \lambda_{p_E} \psi_{ABC}(p'_A, p'_B, p'_C) \\
    \lambda_{p_E} & = \int u(p_A, p_B, -p_A-p_B-p_E, p_E; p'_A, p'_B, p'_C, p'_E)\phi_0(p'_E)dp'_E.
    \end{split}
\end{equation}
Now consider a special case when $|\psi\rangle_{ABC}=|\varphi\rangle_{AB}|\xi\rangle_C$ is a product state, the derivation of (\ref{A3}) is simplified into
\begin{equation}
    \label{A5}
    \begin{split}
    |\psi\rangle_{C,ABE} &= |p=0\rangle_C\otimes \int dp_Edp_Adp_Bdp'_Adp'_Bdp'_C\lambda_{p_E}|p_A\rangle|p_B\rangle\langle p'_A|\langle p'_B|\langle p'_C|\xi\rangle_C|\varphi\rangle_{AB}|p_E\rangle \\
    & = |p=0\rangle_C\otimes\int dp_E \hat{\Gamma}_{p_E} |\varphi\rangle_{AB}|p_E\rangle,
    \end{split}
\end{equation}    
where operator $\hat{\Gamma}_{p_E}$ only acts on particles A and B, defined as
\begin{equation}
    \label{Gamma2}
    \begin{split}
    \hat{\Gamma}_{p_E} & = \int dp_Adp_Bdp'_Adp'_B \times \gamma_{p_E} |p_A\rangle|p_B\rangle\langle p'_A|\langle p'_B|,\\
    \gamma_{p_E} & = \int dp'_C \lambda_{p_E}\xi(p'_C),
    \end{split}
\end{equation}
and $\lambda_{p_E}$ has been defined in (\ref{A4}).

\section{Proof of Eq.(\ref{swapSate2})}
\label{AppendixB}
First we notice that (\ref{state6}) is derived without the assumption of $|\psi\rangle_{ABC}=|\varphi\rangle_{AB}|\xi\rangle_C$. A similar form for $|\varphi\rangle_{AB|C}^m$ can be derived from (\ref{A3}), which is more generic:
\begin{equation}
    \label{B1}
    \begin{split}
    |\varphi\rangle_{AB|C}^m & = \frac{1}{\sqrt{\rho_{m|C}}}\int dp_Adp_B\chi_{p_m}(p_A, p_B)|p_A\rangle|p_B\rangle =\frac{1}{\sqrt{\rho_{m|C}}}|\chi\rangle_{AB|C}^{p_m}, \\
    \rho_{m|C} & ={}_{AB|C}\langle\chi |\chi\rangle_{AB|C}^{p_m}
    \end{split}
\end{equation}
and $\chi_{p_m}$ has been defined in (\ref{A4}). Next step is to evaluate 
\begin{equation}
    \label{B2}
    \begin{split}
    \hat{\cal{S}}_{m}|\chi\rangle_{AB|C}^{p_m} & =  \hat{\cal{P}}_{AC}e^{i\hat{q}_A(\hat{p}_B+p_m)}\int dp_Adp_B\chi_{p_m}(p_A, p_B)|p_A\rangle|p_B\rangle \\
    &=\hat{\cal{P}}_{AC}\int dp_Adp_B\chi_{p_m}(p_A, p_B)|p_A+p_B+p_m\rangle|p_B\rangle.
    \end{split}
\end{equation}
Changing variable $p_C=-p_A-p_B-p_m$ and applying the operator $\hat{\cal{P}}_{AC}$, we get
\begin{equation}
    \label{B3}
    \hat{\cal{S}}_{m}|\chi\rangle_{AB|C}^{p_m} = \int dp_Bdp_C\chi_{p_m}(-p_B-p_C-p_m, p_B)|p_C\rangle|p_B\rangle
\end{equation}
From the definition in (\ref{A4}), $\chi_{p_m}$ depends on $\lambda_{p_m}$, which in turn depends on matrix element $u$. Explicitly,
\begin{equation}
    \label{B4}
    \begin{split}
    \lambda_{p_m}(-p_B-p_C-p_m, p_B) = \int \phi_0(p'_E)dp'_E u(-p_B-p_C-p_m, p_B, p_C, p_m; p'_A, p'_B, p'_C, p'_E).
    \end{split}
\end{equation}
Compared (\ref{A4}), (\ref{B4}), and the definition of $\kappa_{p_m}$ in (\ref{state6}), one recognizes that $\chi_{p_m}(-p_B-p_C-p_m, p_B)=\kappa_{p_m}(p_B, p_C)$. Inserting this identity to (\ref{B3}), we have
\begin{equation}
    \label{B5}
    \hat{\cal{S}}_{m}|\chi\rangle_{AB|C}^{p_m} = |\kappa\rangle^{p_m}_{BC|A}.
\end{equation}
This implies that the measurement probability
\begin{equation}
    \label{B6}
    \begin{split}
    \rho_{m|A} &= {}_{BC|A}\langle\kappa|\kappa\rangle_{BC|A}^{p_m}={}_{AB|C}\langle\chi|\hat{\cal{S}}_m^\dag\hat{\cal{S}}_m|\chi\rangle_{AB|C}^{p_m}={}_{AB|C}\langle\chi|\chi\rangle_{AB|C}^{p_m} = \rho_{m|C}.
    \end{split}
\end{equation}
Given (\ref{B5}) and (\ref{B6}), and the definitions in (\ref{state3}) and (\ref{state6}), we obtain the desired identity
\begin{equation}
    \label{B7}
    \hat{\cal{S}}_{m}|\varphi\rangle_{AB|C}^{p_m} = |\varphi\rangle^{p_m}_{BC|A}.
\end{equation}
(\ref{B6}) is derived without the assumption of product state condition $|\psi\rangle_{ABC}=|\varphi\rangle_{AB}|\xi\rangle_C$. (\ref{swapSate2}) certainly holds true since it is for the special case with the product state condition.

\section{Completeness of $\{\hat{M}_m\}$}
\label{AppendixC}
Note that index $m$ refers to $p_m$ which is a continuous real number, $p_m\in \cal{R}$ for particle E. Changed notation from $p_m$ to $p_E$, the completeness is
\begin{equation}
\label{C1}
    \int dp_E \hat{M}_{p_E}^\dag\hat{M}_{p_E} = \hat{I}_{AB|C}.
\end{equation}
To verify that, we take the complex conjugate of (\ref{state5}), multiply to itself, and evaluation both sides:
\begin{equation}
    L.H.S. = {}_{ABE|C}\langle\psi|\psi\rangle_{ABE|C} = N_L.
\end{equation}
$N_L$ is simply a constant. The inner product can be constructed from either the reduced Hilbert space ${\cal{H}}_{ABE|C}$, or from ${\cal{H}}^{kin}$ as described in Appendix C of Ref.~\cite{Hoehn2018}. To evaluate the $R.H.S.$, we use the property $\langle p'_E| p_E\rangle = \delta (p_E-p'_E)$ and simplify $R.H.S$ as
\begin{equation}
\begin{split}
    R.H.S. &= \int dp'_E dp_E ({}_{AB|C}\langle \varphi|{}_E\langle p_E|\hat{\Gamma}_{p'_E}^\dag)\hat{\Gamma}_{p_E} |\varphi\rangle_{AB|C}|p_E\rangle_E \\
    &= \int dp_E ({}_{AB|C}\langle \varphi|\hat{\Gamma}_{p_E}^\dag\hat{\Gamma}_{p_E} |\varphi\rangle_{AB|C})\\
    &= {}_{AB|C}\langle\varphi|(\int dp_E\hat{\Gamma}_{p_E}^\dag\hat{\Gamma}_{p_E}) |\varphi\rangle_{AB|C}.
\end{split}
\end{equation}
Equating both sides, we obtain
\begin{equation}
\label{C4}
    N_L= {}_{AB|C}\langle\varphi|(\int dp_E\hat{\Gamma}_{p_E}^\dag\hat{\Gamma}_{p_E}) |\varphi\rangle_{AB|C}
\end{equation}
Since $|\varphi\rangle_{AB|C}$ is an arbitrary state vector, the above equation implies
\begin{equation}
\label{C5}
    \int dp_E\hat{\Gamma}_{p_E}^\dag\hat{\Gamma}_{p_E} = N\times\hat{I}_{AB|C},
\end{equation}
where $N$ is a constant. Inserting (\ref{C5}) into (\ref{C4}), we find
\begin{equation}
    \label{C6}
    N = \frac{N_L}{{}_{AB|C}\langle\varphi|\varphi\rangle_{AB|C}}.
\end{equation}
Defining $\hat{M}_{p_E} = \hat{\Gamma}_{p_E}/\sqrt{N}$, we obtain the identify (\ref{C1}).

\section{Proof of (\ref{Proc27})}
\label{AppendixD}
The unitary operator $\hat{U}$ in ${\cal{H}}^{kin}$ is decomposed as 
\begin{equation}
    \begin{split}
    \label{D1}
    \hat{U} = \sum_{i,j}\int dp_Adp_Bdp_Cdp'_Adp'_Bdp'_Cu(p_A, p_B, p_C, \sigma_i; p'_A, p'_B, p'_C, \sigma_j)|p_A\rangle|p_B\rangle|p_C,\sigma_i\rangle\langle p'_A|\langle p'_B|\langle p'_C, \sigma_j|.
    \end{split}
\end{equation}
Then, from definition of $\hat{\Gamma}_m$, we have
\begin{equation}
    \begin{split}
    \label{D2}
    \hat{\Gamma}_m := \langle\phi_m|\hat{U}|\phi_0\rangle=\int d\bar{p}_Cd\tilde{p}_C \phi_m^*(\bar{p}_C)\phi_0(\tilde{p}_C)\langle\bar{p}_C, \sigma_m|\hat{U}|\tilde{p}_C, \sigma_0\rangle.
    \end{split}
\end{equation}
Inserting (\ref{D1}) into (\ref{D2}), and applying the orthogonal identities (\ref{Ortho}), we obtain
\begin{equation}
    \begin{split}
    \label{D3}
    \hat{\Gamma}_m
    & = \sum_{i,j}\int dp_Adp_Bdp_Cdp'_Adp'_Bdp'_Cd\bar{p}_Cd\tilde{p}_C u(p_A, p_B, p_C, \sigma_i; p'_A, p'_B, p'_C, \sigma_j)\phi_m^*(\bar{p}_C)\phi_0(\tilde{p}_C)\\
    & \times\langle\bar{p}_C, \sigma_m| p_A\rangle|p_B\rangle|p_C,\sigma_m\rangle\langle p'_A|\langle p'_B|\langle p'_C, \sigma_0|\tilde{p}_C, \sigma_0\rangle\\
    & = \int dp_Adp_Bdp'_Adp'_B \gamma_m| p_A\rangle|p_B\rangle \langle p'_A|\langle p'_B|\\
    \gamma_m &=\int d\bar{p}_Cd\tilde{p}_C\phi_m^*(\bar{p}_C)\phi_0(\tilde{p}_C) u(p_A, p_B, \bar{p}_C, \sigma_m; p'_A, p'_B, \tilde{p}_C, \sigma_0).
    \end{split}
\end{equation}
Inserting (\ref{D3}) into (\ref{Proc26}), 
\begin{equation}
\label{D4}
\begin{split}
   |\psi\rangle_{C,AB} &= \hat{T}_C\delta(\hat{P})\sum_m\int dp_Adp_Bdp'_Adp'_B\gamma_m|p_A\rangle|p_B\rangle \langle p'_A|\langle p'_B| \\
   &\times\int d\tilde{p}_Ad\tilde{p}_Bd\tilde{p}_C\varphi(\tilde{p}_A,\tilde{p}_B) \phi_m(\tilde{p}_C)|\tilde{p}_A\rangle|\tilde{p}_B\rangle |\tilde{p}_C, \sigma_m\rangle \\
    &=\hat{T}_C\sum_m\int dp_Adp_Bdp'_Adp'_B\gamma_m \delta(p_A+p_B+\tilde{p}_C)|p_A\rangle|p_B\rangle \langle p'_A|\langle p'_B| \int d\tilde{p}_Ad\tilde{p}_Bd\tilde{p}_C\\ 
    &\times \varphi(\tilde{p}_A,\tilde{p}_B) \phi_m(\tilde{p}_C)|\tilde{p}_A\rangle|\tilde{p}_B\rangle |\tilde{p}_C, \sigma_m\rangle \\
    &=\hat{T}_C\sum_m\int dp_Adp_Bdp'_Adp'_B\gamma_m\phi_m(-p_A-p_B)|-p_A-p_B, \sigma_m\rangle_C \\
    & \times |p_A\rangle|p_B\rangle \langle p'_A|\langle p'_B| \int d\tilde{p}_Ad\tilde{p}_B \varphi(\tilde{p}_A,\tilde{p}_B) |\tilde{p}_A\rangle|\tilde{p}_B\rangle\\
    &=\sum_m |p=0, \sigma_m\rangle_C \otimes \{\int dp_Adp_Bdp'_Adp'_B\gamma_m \phi_m(-p_A-p_B) |p_A\rangle|p_B\rangle \langle p'_A|\langle p'_B|\} |\varphi\rangle_{AB}\\
    &= \sum_m |p=0, \sigma_m\rangle_C \otimes\hat{M}_m|\varphi\rangle_{AB}.
\end{split}
\end{equation}
where $\hat{M}_m$ is a modified version of $\hat{\Gamma}_m$,
\begin{equation}
    \label{D5}
    \begin{split}
    \hat{M}_m = \int dp_Adp_Bdp'_Adp'_B\gamma_m \phi_m(-p_A-p_B)|p_A\rangle|p_B\rangle \langle p'_A|\langle p'_B|.
    \end{split}
\end{equation}

\section{Proof of (\ref{Proc28})}
\label{AppendixE}
The derivation of (\ref{Proc28}) is simliar to (\ref{D4}), except this time we pick particle A as QRF.
\begin{equation}
\label{E1}
\begin{split}
   |\psi\rangle_{A,BC}
    &=\hat{T}_A\sum_m\int dp_Adp_Bdp'_Adp'_B\gamma_m \delta(p_A+p_B+p_C)|p_A\rangle|p_B\rangle \langle p'_A|\langle p'_B| \\
    &\times\int d\tilde{p}_Ad\tilde{p}_Bdp_C\varphi(\tilde{p}_A,\tilde{p}_B) \phi_m(p_C)|\tilde{p}_A\rangle|\tilde{p}_B\rangle |p_C, \sigma_m\rangle \\
    &= \hat{T}_A\sum_m\int dp_Bdp'_Adp'_B\gamma_m |-p_B-p_C\rangle_A|p_B\rangle \langle p'_A|\langle p'_B| \\
    &\times\int d\tilde{p}_Ad\tilde{p}_Bdp_C\varphi(\tilde{p}_A,\tilde{p}_B) \phi_m(p_C)|\tilde{p}_A\rangle|\tilde{p}_B\rangle |p_C, \sigma_m\rangle \\
    &=|p=0\rangle_A\otimes \sum_m\int dp_Bdp'_Adp'_B\gamma_m |p_B\rangle \langle p'_A|\langle p'_B|\\
    &\times\int d\tilde{p}_Ad\tilde{p}_Bdp_C\varphi(\tilde{p}_A,\tilde{p}_B) \phi_m(p_C)|\tilde{p}_A\rangle|\tilde{p}_B\rangle |p_C, \sigma_m\rangle.
\end{split}
\end{equation}
Removing the redundancy state of A by applying $\sqrt{2\pi}\langle q_A=0| \otimes$, we have
\begin{equation}
\label{E2}
\begin{split}
    |\varphi\rangle_{BC|A}&:=\sqrt{2\pi}\langle q_A=0|\psi\rangle_{A,BC}\\
    &=\sum_m\int dp_Bdp_C \phi_m(p_C)|p_B\rangle |p_C, \sigma_m\rangle\int d\tilde{p}_Ad\tilde{p}_B\varphi(\tilde{p}_A,\tilde{p}_B)\gamma_m(-p_B-p_C, p_B; \tilde{p}_A, \tilde{p}_B)\\
    &=\sum_m\int dp_Bdp_C \phi_m(p_C)\chi_m(p_B, p_C)|p_B\rangle |p_C, \sigma_m\rangle,
\end{split}
\end{equation}
where
\begin{equation}
\label{E3}
\begin{split}
    \chi_m(p_B, p_C) &=\int d\tilde{p}_Ad\tilde{p}_B\varphi(\tilde{p}_A,\tilde{p}_B)\gamma_m(-p_B-p_C, p_B; \tilde{p}_A, \tilde{p}_B) = {}_A\langle -p_B-p_C|\langle p_B|\hat{\Gamma}_m|\varphi\rangle_{AB},
\end{split}
\end{equation}
and $\hat{\Gamma}_m$ is defined in (\ref{D3}). (\ref{E2}) gives (\ref{Proc28}).

\section{Proof of (\ref{swapSate3})}
\label{AppendixF}
From the definition of $\hat{M}_m$ in (\ref{D5}), we have
\begin{equation}
    \label{F1}
    \begin{split}
    \hat{\cal{S}}_m\hat{M}_m|\varphi\rangle_{AB|C} &=
    \hat{\cal{P}}^n_{AC}e^{i\hat{q}_A\hat{p}_B}\hat{M}_m|\varphi\rangle_{AB|C} \\
    &=\hat{\cal{P}}^n_{AC}\int dp_Adp_Bdp'_Adp'_B\gamma_m \phi_m(-p_A-p_B) |p_A+p_B\rangle|p_B\rangle \langle p'_A|\langle p'_B|\varphi\rangle_{AB}.
    \end{split}
\end{equation}
Changing variable $p_C=-p_A-p_B$, and noting $\langle p'_A|\langle p'_B|\varphi\rangle_{AB} = \varphi_{AB}(p'_A, p'_B)$, we rewrite (\ref{F1}) as
\begin{equation}
    \label{F2}
    \begin{split}
    \hat{\cal{S}}_m\hat{M}_m|\varphi\rangle_{AB|C} &=\hat{\cal{P}}^m_{AC}\int dp_Cdp_Bdp'_Adp'_B \phi_m(p_C)|-p_C\rangle_A|p_B\rangle_B \varphi_{AB}(p'_A, p'_B)\gamma_m(-p_B-p_C, p_B; p'_A, p'_B)\\
    &=\int dp_Cdp_Bdp'_Adp'_B\phi_m(p_C)\gamma_m(-p_B-p_C, p_B; p'_A, p'_B)|p_C, \sigma_m\rangle_C|p_B\rangle_B \varphi_{AB}(p'_A, p'_B)\\
    &=\int dp_Bdp_C \phi_m(p_C)\chi_m(p_B, p_C)|p_B\rangle |p_C, \sigma_m\rangle\\
    &= |\varphi\rangle_{BC|A}^m.
    \end{split}
\end{equation}

\end{document}